\documentclass[twocolumn,showpacs,preprintnumbers,amsmath,amssymb]{revtex4}

\usepackage{graphicx}

\begin{document}

\title
{Spin-polarized stable phases of the 2-D electron fluid
at finite temperatures.
}

\author
{
 M.W.C. Dharma-wardana\cite{byline1}}
\author{Fran\c{c}ois Perrot$^{\ddag}$}
\affiliation{
National Research Council of Canada, Ottawa,Canada. K1A 0R6\\
}
\date{\today}
\begin{abstract}
The Helmholtz free energy $F$ of the interacting 2-D electron fluid is
calculated nonperturbatively using a mapping of the quantum
fluid to a classical Coulomb fluid 
$\lbrack$Phys. Rev. Letters, {\bf 87}, 206 (2002)$\rbrack$.
 For density parameters $r_s$ such that 
$ \sim25 < r_s $, the fluid is unpolarized at all temperatures $t=T/E_F$,
where $E_F$ is the Fermi energy.
For lower densities, the system becomes fully
spin  polarized for $t$ smaller than $\sim 0.35$, and partially
polarized for $\sim 0.35 < t < 2$,
 depending on the density.
At $r_s\sim 25-30$, and $t\sim 0.35$, an ''ambispin'' phase where
$F$ is almost independent of the spin polarization is found. 
These results support recent claims, based on quantum Monte Carlo
results, for a stable, fully spin-polarized phase
at $T$ = 0 for $r_s$  larger than $ \sim 25-26$.
\end{abstract}
%
\maketitle
%
{\it Introduction.}
The physics of the uniform two-dimensional electron
 gas (2DEG)
 depends crucially on the  ``coupling parameter''
$\Gamma$ = (potential energy)/(kinetic energy) arising from the Coulomb
interactions.
The $\Gamma$ for the 2DEG at $T=0$ and mean density $n$
is equal to the mean-disk radius $r_s=(\pi n)^{-1/2}$ per electron.
 In the absence of a magnetic
field or disorder effects, the density parameter $r_s$, the
spin polarization $\zeta$ and the temperature
$T$ are the only relevant physical variables.
As the coupling constant increases, the 2DEG becomes more like
a ``liquid'', but here we continue to loosely
call it an electron ``gas''.
A knowledge of the phase diagram at arbitrary $r_s$,
$\zeta$ and $T$ is of crucial importance to a clear understanding
of many intriguing phenomena (metal-insulator transition\cite{revmod}, 
2-D electrons in CuO layers,
spin-dependent processes etc.)
associated with the 2DEG.

	A number of recent studies\cite{varsen,atta,prl2}
 have suggested that the 2DEG
at $T$ =0 becomes fully spin-polarized for $r_s > r_c$ with
 $r_c\sim 25-30$. However the energy difference
 between the unpolarized and polarized
phases is very close to the statistical uncertainty of the methods used. 
We have completed a study of this transition at finite temperatures
where the energy differences for a larger data base
 become more significant and reveal
a rich phase diagram  which is fully consistent with the
claimed $T=0$ polarized-fluid regime. It turns out that the partially
degenerate strongly correlated 2DEG supports 
partially spin-polarized
phases. This is particularly interesting since the
 3DEG has also been recently shown to have
partially spin-polarized
stable phases\cite{ichi3dphase}
 
 The finite-$T$ phase diagram is examined using  a
computationally simple, conceptually novel
method for calculating the quantum-PDFs and other 
properties (e.g, static response). The method has been
applied to the 2DEG\cite{prl2}, the 3DEG\cite{prl1,prb00}, and
 hydrogen fluids\cite{hyd}, at
arbitrary coupling, spin polarization and 
temperature.
It is
based on identifying  
a ``quantum temperature'' $T_q$ such that the correlation
energy of the  corresponding {\it classical} Coulomb fluid at $T_q$ is
equal to that of the quantum fluid at $T=0$.
After that, hypernetted-chain (HNC)  methods are used for the
classical fluid.
This  classical mapping of
 quantum fluids within the HNC
was named  the CHNC.
The application of the CHNC method to the 2DEG 
required the inclusion of short-range correlation effects 
(``bridge terms'') going  beyond the usual
 HNC approximation. The resulting CHNC procedure for the 2DEG proved
to be remarkably accurate\cite{prl2}.
In fact even an application of CHNC at $T$ = 0, and {\it without}
 bridge corrections
was examined by Bulutay and Tanatar and found to be very useful\cite{balutay}.

We use Hartree atomic units, and consider a unit area $A$ 
with $n$ particles per unit area, with  $n$ = $1/(\pi r_s^2)$. 
The number density of each spin
species is $n_i=n(1+\sigma_i\zeta)/2$,
 where i=1,2 would indicate up or, down spins
with $\sigma_i=\pm1$.
The spin polarization $\zeta$ = $(n_1-n_2)/n$.
The total Helmholtz free energy per unit area, $F$,
has the form $F_0+F_x+F_c$, where $F_0$, $F_x$, $F_c$ are the non-interacting,
first-order exchange, and correlation, contributions to $F$. Thus all
effects beyond first-order exchange are lumped into $F_c$, a convention
used in density functional theory (DFT). The
DFT exchange-correlation
free energy would hence be $F_{xc}$ = $F_x+F_c$.
In indicating spin-components we use the notation $F_i^0, F_i^x$, and $F_i^c$
etc. The canonical ensemble is used so that the Helmholtz free energy
is directly determined, unlike in the approach where the grand-potential
$\Omega$ is determined in terms of the chemical potential which has
to be inverted to obtain the density. This means $F_0$ and $F_x$ (being 
the first-order exchange) are
 determined in terms of the non-interacting chemical potential $\mu_0$,
while $F_c$ has to be determined via a coupling constant integration,
as discussed by us in ref~\cite{prb00,pdw84}
 and in Dundrea et al\cite{dundrea}.

{\it The non-interacting free energy.}
 Isihara and others have considered the expansion of the free energy
about $T=0$ and obtained the leading terms in $r_s$, 
 and $T$ \cite{isihara}.
However, the thermodynamic functions near $T=0$ have a particularly
unfavourable structure for constructing such expansions.
 Since our calculations are nonperturbative,
we evaluate $F_0$ etc., numerically, without expansion. 

Using the the relation, $F_0=E_0-TS_0$, where $E_0$ and $S_0$ are the
 noninteracting internal energy and the entropy, 
$F_0$ can be calculated. An independent evaluation is  obtained from the
relation $F_0=n\mu_0-E_0$.  Also, at $T=0$, we have, in Hartrees
\begin{equation}
F_i^0/n=E_i^0/n = (1+\sigma_i\zeta)^2/4r_s^2
\end{equation}
 Defining the Fermi function $f(e)=1/(1+exp(\beta e-\eta))$
where $\beta=1/T$, $\eta=\mu^0/T$,
\begin{eqnarray}
\label{fzero}
n_i&=&(T/2\pi)I_0(\eta_i)\\
E_i&=&(T^2/pi)I_1(\eta_i)\\
F_i^0&=&n_i\mu_i^0-E_i^0
\end{eqnarray}
where the Fermi integral $I_{\nu}(z)$ is defined as
\begin{equation}
I_\nu(z)=\int_0^\infty\frac{dx x^\nu}{1+e^{x-z}}
\end{equation}
The $I_0$ integral can be evaluated explicitly to give:
\begin{eqnarray}
\eta_i& =&log[exp(1/t_i)-1],\,\,t_i=E_i^F/T\\
E_i^F&=&E_F(1+\sigma_i\zeta),\,\, E_F=\pi n\\
\end{eqnarray} 
The $I_1$ integral can be evaluated directly or reexpressed in
terms of the dilogarithm function.
We have evaluated $F_0$ by several methods;
the dilog function was evaluated using the 
 CERN library routine.

{\it The exchange contribution to the free energy.}
 The first-order unscreened exchange free energy $F_x$ consists
of  $F_x^i$, where $i$ denotes the
 spin species. At $T=0$ these reduce to the exchange
energies:
\begin{equation}
E_i^x/n=-\frac{8}{3\surd{\pi}}
n_i^{1/2}
\end{equation}
Hence,
\begin{equation}
E_x/n=(E_1^x+E_2^x)/n=-\frac{8}{3\pi r_s}[x_1^{3/2}+x_2^{3/2}]
\end{equation}
where $x_1$ and $x_2$ are the fractional compositions
$(1\pm\zeta)/2$ of the two spin species.
At finite $T$, we have:
\begin{equation}
\label{exF}
F_i^x/E_i^x=\frac{3}{16}t_i^{3/2}\int_{-\infty}^{\eta_i}
\frac{I^2_{-1/2}(u)du}{(\eta_i-u)^{1/2}}
\end{equation}
Here $I_{-1/2}$ is the Fermi integral.
The total exchange free energy is $F_x=\Sigma F_i^x$.
The accurate numerical evaluation of Eq.~\ref{exF} requires the removal of 
the square-root singularity by adding and subtracting, e.g,
$I^2(-|\eta|)/(v-|eta|)^{1/2}$ for the case where $\eta$ is negative, and
$v=u$, and so on.

  A real-space formulation of $F_x$ = $F_1^x+F_2^x$
using the zeroth-order PDFs fits  naturally with
the CHNC approach. Thus
\begin{equation}
\label{rspacefx}
F_x/n=n\int \frac{2\pi r dr}{r}\sum_{i<j}h^0_{ij}(r)
\end{equation}
Here  $h^0_{ij}(r)=g^0_{ij}(r)-1$. 
In the non-interacting system at temperature $T$, 
 the antiparallel $h^0_{12}$, viz.,
 $g_{12}^0(r,T)-1$,
 is zero while
$$h_{11}^0({\bf r})
=-\frac{1}{n_i^2}\Sigma_{{\bf k}_1,{\bf k}_2}n(k_1)n(k_2)
e^{i({\bf k}_1-{\bf k}_2){\bf \cdot}{\bf r}} \,\,
=\, -[f(r)]^2$$
Here {\bf k},{\bf r} are 2-D vectors and $n(k)$ is
 the Fermi occupation number at 
the temperature $T$. At $T=0$ $f(r)=2J_1(k_ir)/k_r$ where
$J_1(x)$ is a Bessel function. We have numerically evaluated the
exchange free energy by {\it both} methods,i.e, Eq.~\ref{exF} and
Eq.~\ref{rspacefx}, as a numerical check.


{\it Correlation free energy.}
 The correlation contribution, $F_c$ is evaluated together
with the exchange contribution $F_x$ via a coupling constant integration
over the distribution functions as follows:
\begin{equation}
\label{lambdaintegration}
F_{xc}/n=\int_0^1d\lambda n\int \frac{2\pi r dr}{r}
\sum_{i<j}h_{ij}(r,\lambda)
\end{equation}
Thus we need the interacting PDFs $g_{ij}(r,\lambda)$ for each
$r_s$, $\zeta$ and $T$ for a number of values (we have used 7 values)
 of the coupling
 constant $\lambda$. The PDFs were evaluated using the CHNC method, as
described in Ref.[\onlinecite{prl2}]. We present a brief summary
of the CHNC method in the context of the present study. 

{\it Brief summary of the CHNC method.}
The essence of the CHNC method is to start from the quantum mechanical
 $g^0(r)$ of the non-interacting problem and build up the interacting
$g(r)$ by classical methods.

The fluid of mean density $n$ contains
two spin species
with concentrations
 $x_i$ = $n_i/n$. 
While $T$ is the physical temperature  of
the 2DEG, we consider a classical fluid of  temperature $T_{cf}$
= $1/\beta_{cf}$.
Since the leading dependence of the energy on temperature is quadratic,
we assume that $T_{cf}$= ${\surd{(T^2+T_q^2)}}$ where $T_q$ is the quantum
correction. This is
clearly valid for $T=0$ and for  high $T$, and was justified
in more detail in ref.~\onlinecite{prb00}.
The ``quantum temperature'' $T_q$ was shown to be given by\cite{prl2},
\begin{equation}
\label{2dmap}
t=T_q/T_F=2/[1+0.86413(r_s^{1/6}-1)^2]
\end{equation}
where the Fermi temperature $T_F$ is simply the Fermi energy $E_F=1/r_s^2$
in our units. In effect
 $T_q$ is  really a single-parameter representation
of the density-functional 2D correlation energy. 

The HNC equations for the PDFs of a classical
fluid, and the Ornstein-Zernike(OZ) relations are\cite{hncref}:
\begin{eqnarray}
\label{hnc1}
g_{ij}(r)&=&exp[-\beta_{cf} \phi_{ij}(r)
+h_{ij}(r)-c_{ij}(r) + B_{ij}(r)]\nonumber\\
 h_{ij}(r) &=& c_{ij}(r)+
\Sigma_s n_s\int d{\bf r}'h_{i,s}
(|{\bf r}-{\bf r}'|)c_{s,j}({\bf r}')
\label{hnc2}
\end{eqnarray}
This equation involves two quantities, viz., (i) the
pair-potential $\phi_{ij}(r)$, (ii) the bridge function
$ B_{ij}(r)$\cite{rosen,yr2d}. The other terms, e.g, $c(r)$, is the
 ``direct correlation function''
of the OZ equations, while $h_{ij}(r)$ = $g_{ij}(r)$-1. 

{\it Pair-potential.}
The $\phi_{ij}(r)$ is the pair potential between the
species $i,j$. For two electrons this is
just the Coulomb potential $V_{cou}(r)$.
If the spins are parallel, the Pauli
principle prevents  occupation of the same spatial orbital.
This effect is equivalent to  a
``Pauli exclusion potential'', ${\cal P}(r)$  which when used in the
HNC equation reproduces the $g^0_{ij}(r)$.
Thus $\phi_{ij}(r)$ becomes ${\cal P}(r)\delta_{ij}+V_{cou}(r)$.
The Coulomb potential $V_{cou}(r)$
for a pair of point-charge electrons is  $1/r$.
However,  an electron at
the temperature $T_{cf}$ is 
localized to within a thermal wavelength. Thus,  
we use a ``diffraction
corrected'' form 
\begin{equation}
\label{potd}
V_{cou}(r)=(1/r)[1-e^{-rk_{th}}]
\end{equation}
In the case of the 2DEG 
$k_{th}$ was explicitely determined by numerically
 solving the Schrodinger equation
 for a pair
of 2-D electrons in the potential $1/r$ and calculating the electron density
in each normalized state\cite{pdwkth}. That is, by solving the 2-D
Schrodinger equation
(only $s\,$-waves are needed for the $r\to 0$ case):
\begin{equation}
\frac{d^2R}{dr^2}+\frac{1}{r}\frac{dR}{dr}+2m^*(\frac{k^2}{2}
-\frac{1}{r})R=0
\end{equation}
Here $m^*$ = 1/2 is the effective mass of the electron pair.
The radial function has the asymptotic form:
\begin{equation}
R(r,k)\to [\frac{2}{\pi kr}]^{1/2}cos(kr-\frac{m^*}{k}ln(r)+\delta_{2d})
\end{equation}
The ``on-top'' density $n(0)$ for electrons at the temperature $1/\beta$
is:
\begin{equation}
\label{distr}
\frac{n(0)}{n}=\frac{\int^{\infty}_0 k^2 dk exp(-\beta k^2/2m^*)R^2(0,k)}
{\int^{\infty}_0 k^d dk exp(-\beta k^2/2m^*)}
\end{equation}
Using the same form for the diffraction
correction, i.e., $V_d(r)$=exp$(-k_{th}r)/r$,
it was found that
 $$k_{th}/k^0_{th}=1.158T_{cf}^{0.103}$$
 where $T_{cf}$ is in a.u., and $k^0_{th}$ is the de Broglie
thermal momentum $(2\pi m^*T_{cf})^{1/2}$.

{\it Bridge contribution.}
 The ``bridge'' term  arises from higher-order cluster interactions.
They seem to play a role similar to the ``back-flow corrections''
used in QMC trial wavefunctions\cite{kwon}.
Since the Pauli exclusion already acts to reduce clustering of
parallel spin electrons, the $B_{ii}$ bridge corrections are unimportant.
In Ref.~\cite{prl2} we included the $B_{12}$ terms via
 a simple hard-disk bridge function,\cite{yr2d} with a hard-disk packing 
fraction $\eta_p=0.1175r_s^{1/3}(T_{cf}/E_F)$. Since
$t=T_{cf}/E_F$ is itself a function of $r_s$, the zero temperature
packing fraction $\eta_p(T=0)$ is just a function of $r_s$.
In generalizing the above expression to finite $T$, we note that
$r_s$ is simply the coupling constant $\gamma=P.E/K.E$ per electron.
We take the potential energy $P.E$ = $1/r_s$ for a {\t pair}
 of electrons, and
evaluate the kinetic energy ($K.E.$) from the non-interacting internal energy
as in Eq.~\ref{fzero}.
 Then the finite temperature hard-disk packing fraction is
given by
\begin{equation}
\eta_p=0.235\gamma^{1/3}/[1+0.86413(\gamma^{1/6}-1)^2]
\end{equation}

	The next step is to use the  
$\phi_{ij}(r)$,   
 and solve the coupled HNC+Bridge and OZ equations for the 
two-component {\it interacting}$\,$ fluid. Evaluation
at several values of the coupling constant and integration as in
Eq.~\ref{lambdaintegration} leads to the correlation free energy.

\begin{figure}
\includegraphics*[width=8.0cm, height=8.0cm]{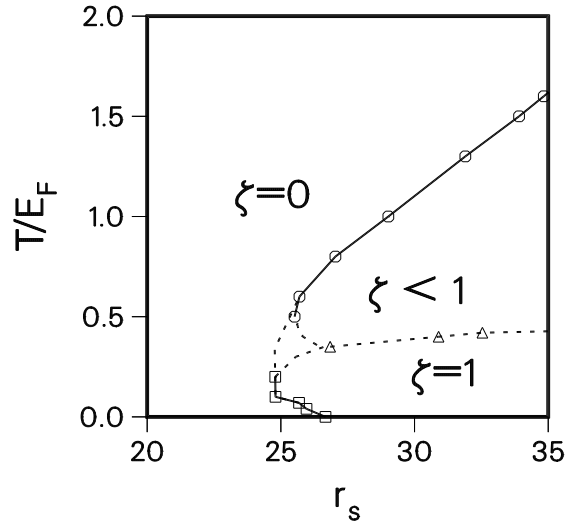}
\caption
{The  phase diagram of the 2DEG as a function of the
density parameter $r_s$ and temperature, showing fully and
partially spin-polarized stable phases.
 The region
near the dotted lines near $r_s\sim 25$ and $t\sim 0.35$ represents
an ''ambispin'' phase where the free energy is almost independent of
the spin polarization.
}
\label{phd}
\end{figure}

{\it Results.}
 The fluid-phase with the lowest Helmholtz free
energy $F(r_s,T,\zeta)$ is the stable phase.
 We have evaluated $F$ for the range 
$t$ =$T/E_F<$ 2, and 5  $< r_s < 40$,
and usually for five values of $\zeta$ and found that
the stable phase is
the unpolarized (paramagnetic) phase for higher temperatures and 
densities.
The initial suggestion of
Varsano and Senatore that the 2DEG at $T$=0
is spin polarized near $r_s\sim 30$, and the
more recent discussion by Attaccalite et al., that a transition occurs at
$r_s$=25.7 are greatly strengthened by our results at finite-$T$. At $T$=0
we find a $\zeta=0\to\zeta=1$ transition at $r_s=r_c(\zeta: 0\to1)=26.7\pm 0.8$,
 but the stabilization energy
is within the inherent uncertainities of the CHNC method. However,
when taken together with all the results for neighbouring finite-$T$
values, a  more plausible picture emerges ( Fig.~\ref{phd}).
When the
temperature is increased, the value of $r_c(\zeta: 0\to1)$ decreases somewhat
and then begins to increase. However, when $t$ reaches $\sim 0.35$, 
partially polarized states with $\zeta \sim 0.7$ to 0.9 begin to
compete for stability. This partially polarized state remains 
stable as long as the temperature is sufficiently low. 
An interesting aspect of the partially polarized regime is the existence
of ``ambi-spin'' regions where the spin projection is NOT
 unambiguously determined
since the energy differences as a function of the polarization
are very small ( see below). For $t > \sim 1.5$,
the stable phase becomes the usual unpolarized (paramagnetic) phase.

\begin {figure}
\includegraphics*[width=8.0cm, height=8.0cm]{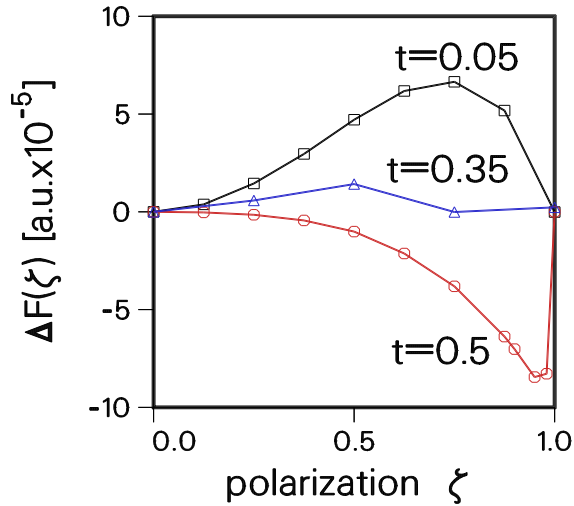}
\caption
{
The polarization dependence of the Helmholtz free energy plotted as
$\Delta F(\zeta)=F(r_s,t,\zeta)-F(r_s,t,0)$ at several temperatures
and densities close to a tranition (see the text).
}
\label{fz}
\end{figure}

A plot of the energy difference $\Delta F(r_s,t,\zeta)= F(\zeta)-F(0)$,
as a function of the
spin polarization is shown in Fig.~\ref{fz} at several
temperatures and $r_s$ values close to one of the transition
lines. Thus the $t=0.05$ curve is at  $r_s\sim 25.86$ which corresponds to 
$r_c(\zeta: 0\to1)$.
  This curve is similar to that
given in Fig.~1 of Ref.~\cite{atta}. However, the the transition regions
near $r_s\sim 26$ and $t\sim 0.35$ provides an example of a phase where
the Helmholtz fee energy is almost independent of the spin polarization.
Such an ``ambispin phase'' would be very sensitive to any external
effects (impurities etc.,) that we have not considered in this study.
The regions in Fig.~\ref{phd} close to the dotted line, begining from the
''triangular'' region near $r_s\sim 25$ and $t\sim 0.35$ are such regions.
The partially spin polarized state is quite unusual since, in
some regions, a very light deviation from $\zeta=1$ leads to a
significantly large stability.  Thus the curve at $t=0.5$  is typical of
densities near $r_s\sim 30-32 $ where the partially polarized stable is
very close to unity but differs slightly from unity.

At $T=0$ the liquid phase becomes a Wigner crystal near
$r_s \sim 35$. However, as the temperature is increased, this value will
also increase. At low temperatures the crystal would be in equilibrium with
the fully spin polarized phase. We have not investigated the equilibrium
between the solid and liquid phases.

Thus the finite-$T$ 2DEG system  provides many intriguing
questions while complementing our understanding of the
$T=0$ system.
Our computer codes 
for generating the PDFs, the $f_{xc}((r_s,T,\zeta)$ and related properties
of the 2D electron fluid may prove to be useful for
further studies of this system. The codes may be accessed via
 the internet\cite{web}.

	In conclusion, we have evaluated the total Helmholtz free energy
$F(r_s,T,\zeta)$ = $F_0 + F_x + F_c$, in the canonical ensemble,
 for a range of $r_s <$ 40, $t\le 2$,
and $0\le z \le 1$. 
 The terms $F_0$ and $F_x$ were
evaluated by several methods, without resorting to perturbation
expansions, while the
$F_c$ was evaluated via a coupling constant integration of the 
pair-distribution functions obtained from the CHNC equations. The finite-$T$
results are consistent with recent claims, based on
quantum Monte Carlo calculations,
 of the existence of a fully spin-polarized stable
phase at $T$ =0 and $r_s\sim 25-26$. 
Our results show that the finite-$T$ phase diagram of
 the clean, uniform 2DEG
contains stable fully and partially spin-polarized phases.
Ambispin phases where the free energy is quasi-independent of the spin
polarization are also found.
It is hoped that these results would
stimulate new experiments and finite-$T$ QMC simulations.

%
%

%

%

\end{document}